# Ultrafast and broadband tuning of resonant optical nanostructures using phase-change materials

Miquel Rudé[1*], Vahagn Mkhitaryan[1*], Arif E. Cetin[2,4], Timothy A. Miller[1], Albert Carrilero[1], Simon Wall[1], F. Javier García de Abajo[1,3], Hatice Altug[2], Valerio Pruneri[1,3†]

1 ICFO - Institut de Ciències Fotòniques, The Barcelona Institute of Science and Technology, 08860, Castelldefels, Barcelona, Spain

2 Bioengineering Deparment, EPFL – Ecole Polytechnique Federale de Lausanne, 1015, Lausanne, Switzerland

3 ICREA—Institució Catalana de Recerca i Estudis Avançats, 08015 Barcelona, Spain

4 Department of Biological Engineering and Koch Institute for Integrative Cancer Research, Massachusetts Institute of Technology, Cambridge, MA 02139, USA.

## Abstract

The functionalities of a wide range of optical and opto-electronic devices are based on resonance effects and active tuning of the amplitude and wavelength response is often essential. Plasmonic nano-structures are an efficient way to create optical resonances, a prominent example is the extraordinary optical transmission (EOT) through arrays of nano-holes patterned in a metallic film. Tuning of resonances by heating, applying electrical or optical signals has proven to be more elusive, due to the lack of materials that can induce modulation over a broad spectral range and/or at high speeds. Here we show that nano-patterned metals combined with phase change materials (PCMs) can overcome this limitation due to the large change in optical constants which can be induced thermally or on an ultrafast timescale. We demonstrate resonance wavelength shifts as large as 385 nm --an order of magnitude higher than previously reported-- by combining properly designed Au EOT nanostructures with $Ge_2Sb_2Te_5$ (GST). Moreover, we show, through pump–probe measurements, repeatable and reversible, large-amplitude modulations in the resonances, especially at telecommunication wavelengths, over ps time scales and at powers far below those needed to produce a permanent phase transition. Our findings open a pathway to the design of hybrid metal-PCM nanostructures with ultrafast and widely tuneable resonance responses, which hold potential impact on active nano-photonic devices such as tuneable optical filters, smart windows, bio-sensors and reconfigurable memories.

---

[*]These authors made equal contribution
[†]valerio.pruneri@icfo.eu

Nanophotonic devices incorporating metallic elements can support plasmons, which are collective oscillations of conduction band electrons driven by an external electromagnetic field[1]. Plasmons can confine and guide light well below the diffraction limit, and when supported by suitably engineered nanostructures, they enable the design of disruptive devices for a wide range of applications, including perfect lenses[2], biosensors[3], modulators[4], plasmonic tuning[5] and integrated waveguide circuits[6]. Plasmons also play an important role in the phenomenon of extraordinary optical transmission (EOT) of visible and infrared light through periodic arrays of subwavelength nanoholes drilled in metallic films. The observation of transmission resonances in these arrays is attributed to the resonant interaction between holes mediated by surface plasmons propagating on the film surfaces[7]. More precisely, transmission peaks emerge close to the Wood anomalies[8] and are well explained in terms of geometrical resonances in the periodic lattice[9,10,11].

An important challenge in the design of plasmonic nanostructures is the precise control of their optical responses in order to meet the requirements of specific device applications. This can be accomplished by casting nanostructures with appropriate materials and geometries. However, such an approach is static and limited by material inhomogeneity and fabrication tolerances. More critically, many applications (e.g., optical switching and modulation) rely on the ability to actively change the spectral phase and amplitude of the optical response. Dynamic tuning could provide a way to reach a stable working point and at the same time modulate the response around it. To this end, the high sensitivity of surface plasmons to local variations in the surrounding refractive index has been exploited in several approaches to achieve dynamical tuning by using active materials whose refractive index is actively changed by external stimuli. In particular, modulation of surface plasmon polaritons (SPPs) has been demonstrated using electro-optic or other effects[4]. For devices based upon EOT, liquid crystals have been used to produce controllable wavelength shifts of 10-20 nm in the near-infrared (NIR)[12]. However, larger tuning ranges and faster modulation than those offered by these materials are needed for a wide range of applications[13].

Phase-change materials (PCMs), such as $Ge_2Sb_2Te_5$ (GST), have two stable phases with very large contrast in their optical and electrical properties[14]. When PCM and metallic nanostructures supporting SPPs are properly coupled, the large changes in optical constants associated to the metal-insulator transition can be used to efficiently tune resonances. Phase transitions in PCMs can be externally driven by varying the temperature or by applying electrical or optical pulses. Importantly, these transitions can occur on the nanosecond timescale[15,16], while the resulting phase is remarkably stable for several years[16]. Besides the widespread use of PCMs in optical storage media[14] and phase-change random access memories (PCRAM)[17], these properties also make them suitable for nanophotonics, where several applications have already been demonstrated[18,19,20,21]. Additionally, the capability to retain their state after a phase transition offers an opportunity to design novel reconfigurable optical elements.

In this work, we explore new designs for combining thin films of GST with arrays of nanoholes patterned in Au films and demonstrate unprecedented ultrafast and broadband optical tuning in the visible and near-infrared (vis-NIR) spectral regions of EOT devices. These coupled EOT structures provide optical resonances over a large wavelength range, while the GST permits the tuning that metals are lacking, which would not possible only using Au nano-holes. More specifically the aim of this work is twofold: we show large wavelength and modulation tuning of the EOT resonances based on phase (structural) transitions of GST using thermal- and current-induced effects; additionally we demonstrate ultrafast and large amplitude modulation of the EOT resonances based on changes of the resonant bond polarizability of GST[22]. Importantly, the phase-transition-induced resonance change is permanent and remains after the temperature (current) signal is switched off, while the ultrafast optical changes are completely reversible for pump powers below the threshold value that would induce phase transitions[22]. We show that the GST optical properties modify the SPPs of the Au perforated film, with the latter being essential for the existence of plasmonic resonances. The combination of temperature (current) and ultrafast-optical-pumping induced changes thus enlarges the range of wavelength, amplitude, reversibility and time operation in EOT-based tunable optical devices.

As summarized in Table 1, we study three different EOT geometries (samples A, B, and C, see below) in which we achieve a significant degree of vis-NIR optical tuning by inducing transitions of a GST layer through thermal, electrical, or optical control signals. We demonstrate spectral shifts in the resonant wavelength as large as 385 nm, combined with modulation depths of >60% associated with phase (structural) transition by applying temperature and electrical currents. We additionally show ultrafast sub-picosecond dynamics by optical pumping, yielding >30% changes in transmission in the absence of any phase (structural) transition. The range, modulation and speed of wavelength tuning of the investigated EOT structures are over one order of magnitude larger than those previously reported, opening new possibilities for nanophotonic devices containing PCMs, especially when one considers that electrical (thermal) and optical tuning mechanisms can be simultaneously applied.

To measure the extent to which the addition of a GST layer can alter the EOT of a patterned gold film, sample A was made. Sample A consists of a hexagonal nanohole array patterned on a 40 nm thick Au film (Fig. 1b). A 20 nm GST film is deposited both on top of the Au film and inside the nanoholes. The device is fabricated on a fused SiO$_2$ substrate using colloidal lithography to create an array of closely-packed polystyrene beads (PS beads) with a nominal diameter of 470 nm, followed by reactive-ion etching (RIE) to shrink the diameter down to 250 nm. Finally, after 3 nm adhesive Ti layer a 40 nm thick Au film is deposited on top using a thermal evaporator, and, after removal of the PS beads, a capping layer of 20 nm amorphous GST is deposited using RF sputtering (see Methods). The result is a metal film pierced by a hexagonal array (period a = 470 nm) of nanoholes (diameter D = 250 nm) and coated on top with an amorphous GST film - see scanning electron microscopy (SEM) image (Fig. 1 b of SI).

The optical transmission spectrum for sample A is shown in Figure 1a for wavelengths in the 400-2400 nm range. The transmission of a gold nanohole array without GST is also shown as a reference (orange curve). A peak in transmission at ~500 nm, associated with EOT, can be clearly identified in all three cases. The transition between amorphous (blue curve) and crystalline (green curve) phases of GST is triggered by heating the sample on a hot plate at 200 ºC. Shifts in the resonance wavelength (from 1330 to 1715 nm) are indicative of this phase transition in the GST film. Heating for about 1 minute is sufficient to completely crystallize the film, and no further changes in the spectrum are observed after subsequent thermal treatments (see SI, Section I). After crystallization, we observe a 385 nm redshift in the resonant wavelength, accompanied by >60% decrease in peak transmission. Figures 1b and c (inset) show finite-difference time-domain (FDTD) simulations of the electric field distribution and optical transmission, respectively, for this sample. The simulated transmission agrees well with the experimental results, thus confirming that the large tuning is associated with significant changes in optical constants (refractive index and absorption) of the GST upon thermally driven phase (structural) transition. The calculated electric field distribution supports the idea that the GST modifies the SPPs of the nanopatterned Au.

The same sample A was then used to investigate changes in EOT resonances via optical pumping associated with the ultrafast and reversible dynamics of resonant bond polarizability. For this purpose, a pump-probe experiment was performed (Figure 2a). Starting with the GST in its crystalline phase, the sample was irradiated with pump (800 nm wavelength) and probe pulses at various delay times and wavelengths. For pump fluences below the threshold required to amorphize the sample, it is known that GST is capable of transiently and rapidly acquiring values of the dielectric function close to those of the amorphous state without completing the phase transformation[22]. More importantly, this effect is ultrafast. Figure 2a shows the dynamics of Sample A during the initial 3 ps following the pump pulse. In order to make sure that the dynamics are related to the GST nanopatterning, we further repeat the same pump-probe experiment using a bare Au nanohole sample (i.e., without GST), which shows pumping the Au film does not modify the Au SPPs (no changes in transmission were observed). At telecom wavelengths around 1550 nm (Figure 2b) the combined sample shows a decrease in transmission due to the ultrafast change in the dielectric function of GST. The modulation in transmission is fast, reaching a peak value in 100 fs (resolution limited) and recovering after a few ps. Furthermore, the magnitude of the modulation is already >30% for modest excitation power.

To verify that the GST enables control of the Au SPPs, a sample was made without GST present in the patterned holes. Sample B consists of a nanohole array with the same geometry as in Sample A, but the 20 nm thick GST film is not present inside the holes and only covers the top flat Au surface (see SI section II and Figure S2). This sample is also used to demonstrate electrical tuning of the combined Au EOT/GST structure. The fabrication procedure is identical to Sample A, except that the PS beads are removed after deposition of GST. In this sample, the GST amorphous-to-crystalline transition driven by heating produces a resonance wavelength blue shift ~35 nm, while the transmission slightly decreases from 28% to 24%. These changes are in agreement with our FDTD simulations (see SI, Section II) and confirm the reliability of modeling, which can be used in designing and optimizing future designs. We also demonstrate optical tuning of Sample B by triggering the GST transition with electrical signals. Using two lateral 40 nm thick Au films as planar contact electrodes and applying a DC current (3.5 V, 1.5 A), crystallization is achieved in 20 s due to Joule heating of the Au film underneath the GST. The results (see SI, Section II) confirm that the electrically driven transition to a crystalline phase produces the same optical response as that obtained by direct heating.

Certain applications, such as tunable filters or optical biosensors, require narrower resonances than those achieved through the non-perfectly periodic metal arrays of samples A and B. With this purpose in mind, we investigated periodic nanohole arrays perforated in a 100 nm free-standing $Si_3N_4$ membrane covered with a 125 nm thick Au film and a 10 nm GST film (Fig. 3b). The device is fabricated using a combination of deep ultraviolet lithography and RIE, as described in Ref. 12. This allows us to pattern a square lattice (period a = 600 nm) of nanoholes (diameter D = 200 nm) displaying much sharper resonances in the transmission spectrum (Sample C). Figure 3a shows the measured and simulated transmission spectra for Sample C without GST and with GST in both phases. Note that the wavelength range and the resonance bandwidths are respectively about 20 times and 10 times smaller than the corresponding values for Sample A. Similar to Sample B, crystallization induces a blue shift, except that the effect is now weaker (only 13 nm shift), which is consistent with the smaller thickness used for the GST layer (10 nm).

The physical origin of the large tuning observed in the experiments lies in the optical contrast of GST. After phase transition, both its refractive index and absorption coefficient increase for wavelengths above 450 nm, which lower transmission and shifts the resonant peak in the crystalline phase to either shorter or longer wavelengths, respectively, depending on whether the GST is present only on top of the Au film (samples B and C) or also inside the holes (sample A).

We gain further insight into the effect of GST phase transitions on the SPP-mediated EOT through a qualitative description based on a semi-analytical model that describes the interaction of light with the nano-structured surface[11]. Because the hole diameter and array period are smaller than the wavelength, a single beam contributes to the far-field transmission, which we calculate by treating each hole as an equivalent set of dipoles. More precisely, under normal incidence, surface-parallel magnetic dipoles dominate the response, one on either side of the film[11] (see SI, Section III). Then, the transmission coefficient of the array can be expressed through the induced magnetic dipole moment $m_2$ on the far side of the film as

$$T = \left| \frac{2\pi i k m_2}{A}(1 - r_{23}) \right|^2,$$

where $m_2$ represents the analytical expression

$$m_2 = \frac{\alpha_{M2}}{(1-\alpha_{M1}G_1)(1-\alpha_{M2}G_2)-\alpha'_{M1}\alpha'_{M2}G_1G_2}H^{ext}.$$

Here, $H_{ext}$ is the external magnetic field at the opening of hole, including the specularly reflected field from the metal surface, $\alpha_{M1}$ and $\alpha_{M2}$ ($\alpha'_{M1}$, $\alpha'_{M2}$) are the magnetic polarizabilities of the upper and lower sides of the hole as seen from the near (far) side, and $G_1$ and $G_2$ are the lattice sums over dipole-dipole inter-hole interactions on the near and far sides, respectively. These lattice sums display characteristic divergences which appear as lattice resonances in the spectra (see SI, Section III). The magnetic polarizabilities and lattice sums both depend on the surrounding environment and geometrical parameters and are therefore sensitive to the presence and optical properties of the GST material. Further details of this analytical model are offered in the SI.

Despite its simplicity, this model yields spectra (Figure 1, inset) in qualitative agreement with experiment. In particular, it also describes large shifts in the GST-covered sample compared to bare gold. The model permits ascribing the observed resonances to the (±1,0) and (0,±1) lattice resonances, with the shift from the Wood anomaly condition (i.e., that these diffracted beams become grazing) originating in the interaction with the GST-phase-sensitive environment. Indeed, the phase-change-driven shift of Sample A is much larger than those of Sample B and C because the GST is present inside the hole, thus producing a stronger modification of the hole polarizability.

In contrast to equilibrium regime, the ultrafast tuning is associated with changes in the resonant bond polarizability produced by photo-induced breakdown of the resonant bonds of the crystalline phase[22]. In this intermediate, non-equilibrium state, a large decrease in real and imaginary parts of the dielectric function could be observed similar to the equilibrium amorphous state. This change, in turn, leads to a change in transmission of the structure in the same way as in the equilibrium regime discussed above, since the tuning process does not depend on the actual mechanism used to tune the dielectric function of the environment, but rather on its instantaneous values.

The plasmonic nature of the observed resonances are further supported by FDTD numerical simulations (Figure 1.a, broken curves), which yield spectra in excellent agreement with experiment (solid curves) and allow us to explore the near field under either off- (Figure 1.c) or on-resonance (Figure 1.d) conditions. The off-resonance field is mainly distributed in the near side of the film, with little penetration inside the hole. In contrast, the resonant field is strongly localized on the edges of the hole openings and reaches higher absolute values. These results are consistent with our interpretation of plasmon-mediated lattice resonances, involving large hole polarization at the transmission maxima.

In conclusion, we have demonstrated optical tuning of resonant nanohole array structures patterned in an Au film using the well-known phase-change material GST (Figure 4). The high contrast in the optical properties of GST enables thermal and electrical tuning of the resonant response with spectral shifts as large as 385 nm and modulation depths larger than 60%, well beyond those previously reported for other designs. Moreover, our work shows that the tuning can be optically induced without the GST undergoing any phase transition, thus extending the device lifetime. The resulting optically induced modulation is still large (30 %) in the experiments but can be improved and it occurs over an ultrafast timescale in the picosecond domain.

Although the GST cycleability is $10^5$-$10^7$ cycles[16] and is limited by accumulated stress with the surrounding material due to density differences between the amorphous and crystalline phases (Δρ = 4 %), in this case the absence of a phase transition makes it possible to continuously modulate the response on the ps timescale without any long term permanent change or damage. The ultimate repetition limit of this modulation should be given by thermal diffusion of the pump energy through the substrate before heat accumulation causes an eventual GST amorphization. This is a consideration in devices with a high degree of integration, which could be potentially optimized for heat management (e.g., by resorting to highly thermally conductive materials) to enable high repetition rates in the range of current microelectronics computation speeds. Note that the measurement presented in Figure 2 consists of more than 500,000 excitations without detectible deterioration of the optical response of the material. It is worth pointing out that the gold nano-hole pattern and the GST thin films are both quite uniform despite having used scalable deposition and lithography techniques. This was confirmed by measuring the optical response while scanning the beam across the sample area, which no significant variations observed.

With proper scaling, the proposed designs, which we demonstrate in the visible and near-infrared spectral regions, can be extended to the mid-infrared, as phase change materials also exhibit large changes of optical constants at longer wavelengths. The low-cost nanofabrication methods used in this work[23,24] for patterning nanostructures incorporating phase change materials hold great potential as the basis to manufacture ultrafast and tunable optical devices operating over a wide spectral range.

## Methods

**Fabrication of nanohole arrays using colloidal lithography**

Nanohole arrays were fabricated using colloidal lithography. 50 μL of polystyrene beads (PS) solution (10 % concentration), with a nominal diameter of 470 nm, were mixed in ethanol in a 1:1 volume ratio and placed in an ultrasonic bath for 30 minutes. A laminar flow of the prepared solution was then created on the surface of distilled water using a curved pipette. Water was contained in a small box where the $SiO_2$ substrates had been immersed previously. After a few minutes, the water surface was covered by a polycrystalline monolayer of hexagonally packed PS beads and after removing the distilled water these monolayers were deposited on top of the substrates. Shrinking of the PS beads down to 250 nm was achieved with reactive-ion etching (RIE) using $O_2$ plasma for 4:15 at 100 W. Then, a 5 nm Ti layer and a 40 nm Au layer were thermally evaporated on top. The Ti layer was used as an adhesion layer between the substrate and the Au. Finally, after (sample A) or before (sample B) removing the PS beads using scotch tape, a 20 nm thick layer of GST was deposited by RF co-sputtering from high purity targets of GeTe and $Sb_2Te_3$ in an Ar atmosphere (3.75 mTorr) for 90 s. X-ray diffraction of GST films prepared under the same conditions confirmed the initial phase of the material to be amorphous.

**Transmission measurements**

Transmission measurements for Samples A and B were performed at normal incidence for wavelengths between 300 nm and 2400 nm in a commercial spectrophotometer using a 5 nm wavelength step and a rectangular beam of 3x8 mm$^2$.

Optical characterization of the Au nanohole arrays suspended on a $Si_3N_4$ membrane was performed via spectroscopic measurements, using an unpolarized broadband white light source. Transmitted light from the chip was collected by a high-magnification objective lens (100× Nikon objective lens with NA of 0.6 embedded in a Nikon Eclipse-Ti microscope) coupled with an optical fiber and recorded with a Maya 2000Pro spectrometer.

**Pump-probe experiments**

The time response of Sample A was measured using an optical pump-probe setup. 35 fs laser pulses at 800 nm with a fluence of 5 mJ/cm$^2$ were used to pump the sample with a repetition rate of 80 Hz. Infrared pulses with a duration of 60 fs generated in an optical parametric amplifier measured the transmission of the sample at different time delays from 1150 - 2150 nm in 100 nm spectral steps. The transmitted light was collected using a photodiode and lock-in detection. The amplitude of the diode was recorded as a function of probe delay, generating the signal presented. The reduced repetition rate was needed to avoid both cumulative heating of the sample film by the pump pulses and melting followed by subsequent re-amorphization.

## Acknowledgements

The work has received financial support from the Spanish Ministry of Economy and Competitiveness (MINECO), the "Fondo Europeo de Desarrollo Regional" (FEDER) through grant TEC2013-46168-R by NATO's Public Diplomacy Division in the framework of "Science for Peace".

## Authors Contributions

VP initiated the research and with the help of MR, VM and HA designed the experiments. MR with the help of AC fabricated the samples on fused silica. AEC fabricated the sample on $Si_3N_4$ membrane. MR, VM, AEC performed the experiments, except for the ultrafast measurements performed by TM and SW. VM and JGdA provided the theory. All authors contributed to the interpretation of the results and writing.

## Competing financial interests

The authors declare no competing financial interests.

## Corresponding Authors

valerio.pruneri@icfo.eu

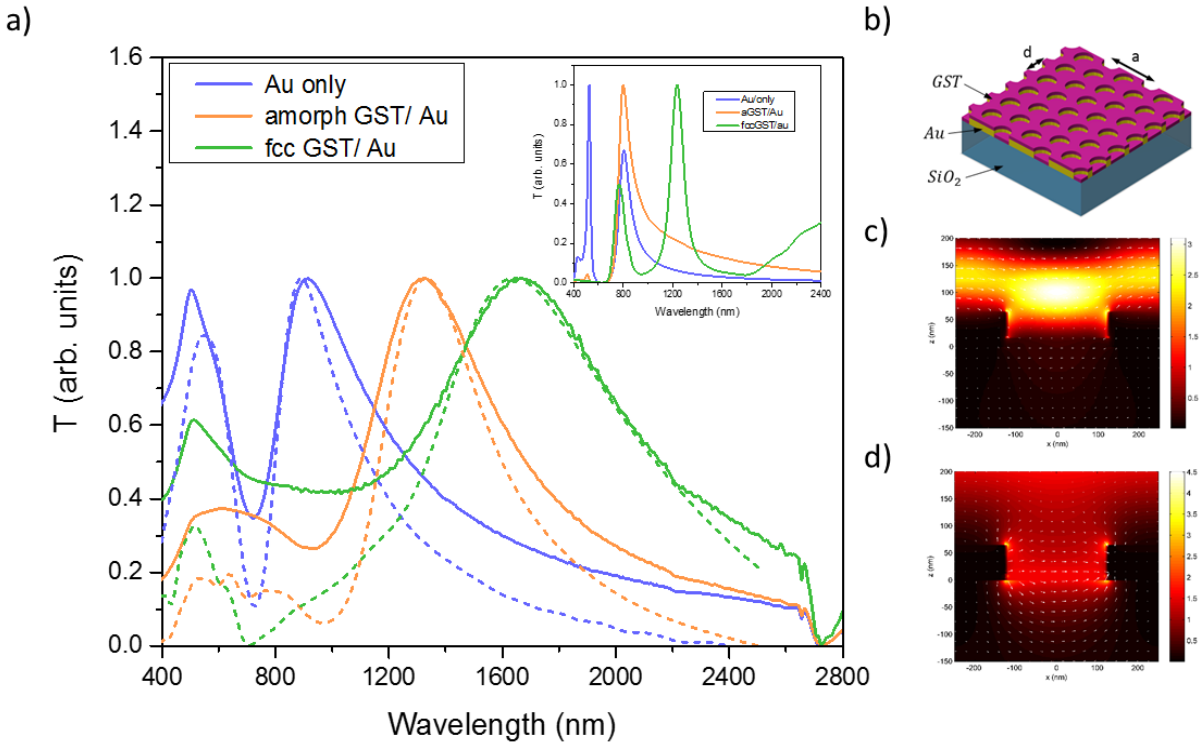

**Figure 1**. **a,** Experimental transmission (solid curves) of sample A with the GST in its amorphous (orange) and crystalline (green) phases, compared with FDTD simulations (broken curves). The transmission of a bare Au nanohole array with the same geometry is also included as a reference (blue). The actual (measured) peak transmission amplitude for the crystalline case is < 40% of that for the amorphous one (this corresponds to >60% modulation depth). Inset: corresponding calculated transmission curves based on an analytical dipole model. **b,** Schematic of sample A consisting of a Au nanohole array covered with a 20 nm thick GST film. **c,d** Electric field distribution under off- (c, $\lambda$ = 1200 nm) and on-resonance (d, $\lambda$ = 1715 nm) conditions when the GST is in the crystalline phase.

| | D | a | GST thickness | Gold Thickness | GST in holes | Thermally tuned | Electrically tuned | Optically modulated | Δλ | Resonance width |
|---|---|---|---|---|---|---|---|---|---|---|
| **Sample A** | 470 nm | 250 nm | 20 nm | 40 nm | Yes | Yes | No | Yes | 385 nm | Broad |
| **Sample B** | 470 nm | 250 nm | 20 nm | 40 nm | No | Yes | Yes | No | 35 nm | Broad |
| **Sample C** | 600 nm | 200 nm | 10 nm | 125 nm | No | Yes | No | No | 13 nm | Narrow |

**Table 1.** Summary of the three configurations used in this work and the different tuning mechanisms.

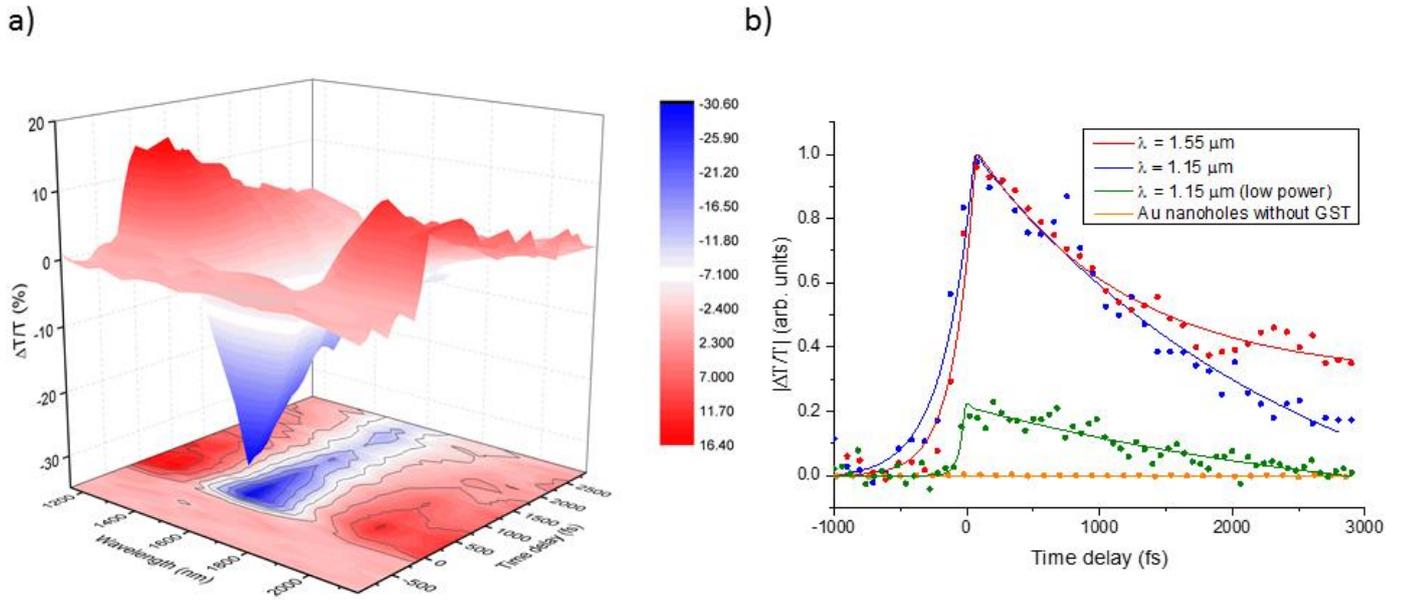

**Figure 2. a,** Time response of Sample A in the λ = 1150 - 2100 nm range during the initial 3 ps. The sample consists of an Au nanohole array covered with a 20 nm thick GST film. **b,** Normalized time response of the device at λ = 1.15 μm and λ = 1.55 μm using a pump fluence of F = 5 mJ /cm$^2$. For both wavelengths the device shows a prompt change in transmission after photoexcitation followed by an exponential recovery. At λ = 1.15 μm and lower pump fluences (F = 1.5 mJ/cm$^2$, green circles) the sample shows the same behaviour with a smaller modulation. The time response of a device containing only a Au nanohole array without GST (orange circles) is also shown at λ = 1.15 μm.

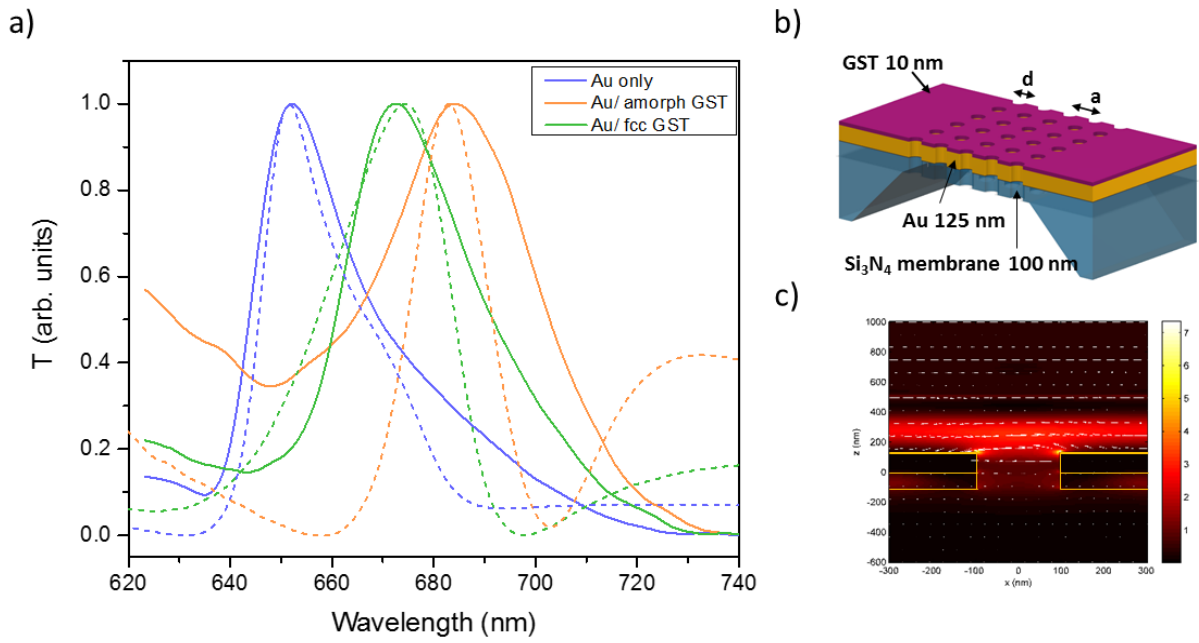

**Figure 3. a,** Normalized transmission (solid curves) in both the amorphous (orange) and crystalline (green) phases, compared with FDTD simulations (broken curves). Transmission of the same device without the GST film on top is also included as a reference (blue). **b,** Schematic of sample C consisting of a Au nanohole array suspended on a $Si_3N_4$ membrane with a top layer of GST (10 nm). **c,** Electric field distribution at the resonance wavelength of the crystalline phase (λ = 670 nm).

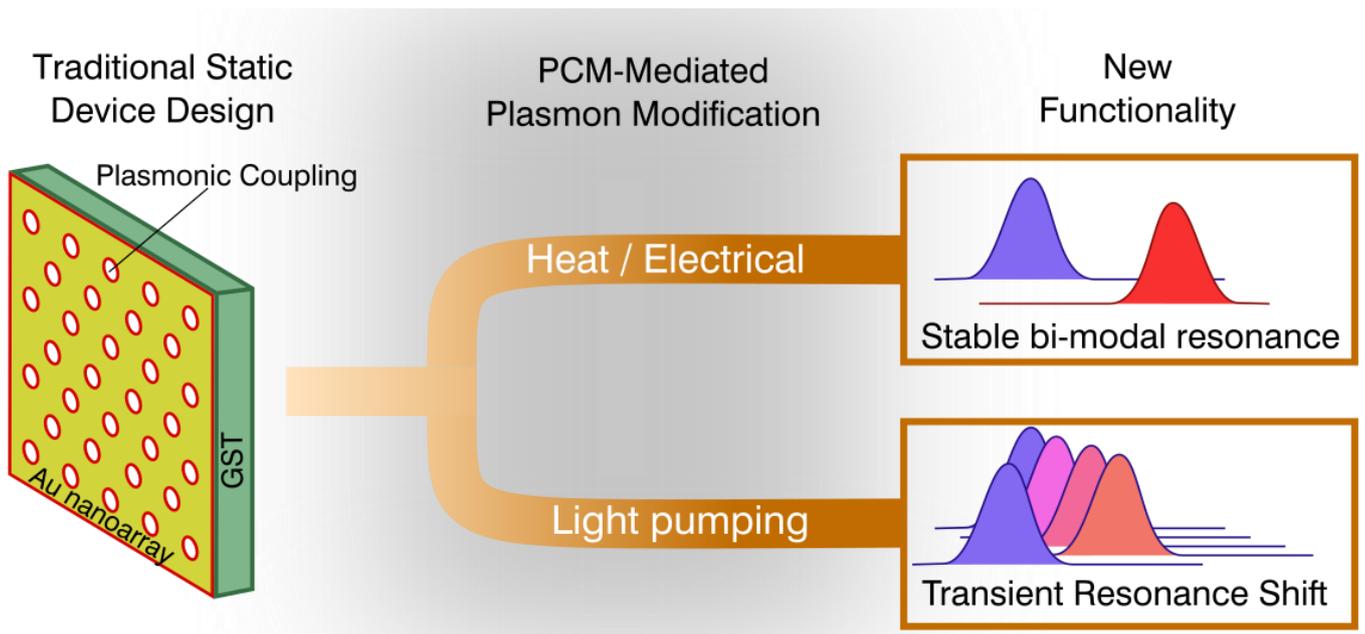

**Figure 4.** Summary of combined GST/EOT Device Control. The addition of a phase change material into traditional EOT device design greatly improves device flexibility. One retains all the control traditionally offered by static EOT device design over resonance wavelength and bandwidth. However, the inclusion of GST allows the EOT device to be controlled in all the ways that one may change GST optical properties. This allows the use of the device in a bi-modal way with long-term stability, an ultrafast way with transient modification, as well as in a configuration directly compatible with existing electro-optic devices.

# SUPPLEMENTARY INFORMATION: ULTRAFAST BROADBAND TUNING OF RESONANT OPTICAL NANOSTRUCTURES USING PHASE-CHANGE MATERIALS

*Miquel Rudé, Vahagn Mkhitaryan, Arif E. Cetin, Timothy A. Miller, Albert Carrilero, Simon Wall, F. Javier García de Abajo, Hatice Altug, Valerio Pruneri*

## I. Thermal treatment of GST

Crystallization of GST is achieved by placing the sample in a hot plate at 200 ºC. Figure 1a shows the evolution of the resonance both for the system consisting of GST and Au nanoholes and the system consisting only of Au nanoholes. The first point represents the system before any thermal treatment. After 1 minute (2$^{nd}$ point) the resonance does not shift anymore, indicating that the GST layer is completely crystallized. Figure 1b shows the SEM image of one of the samples used in this work, consisting of a Au nanohole array covered with GST.

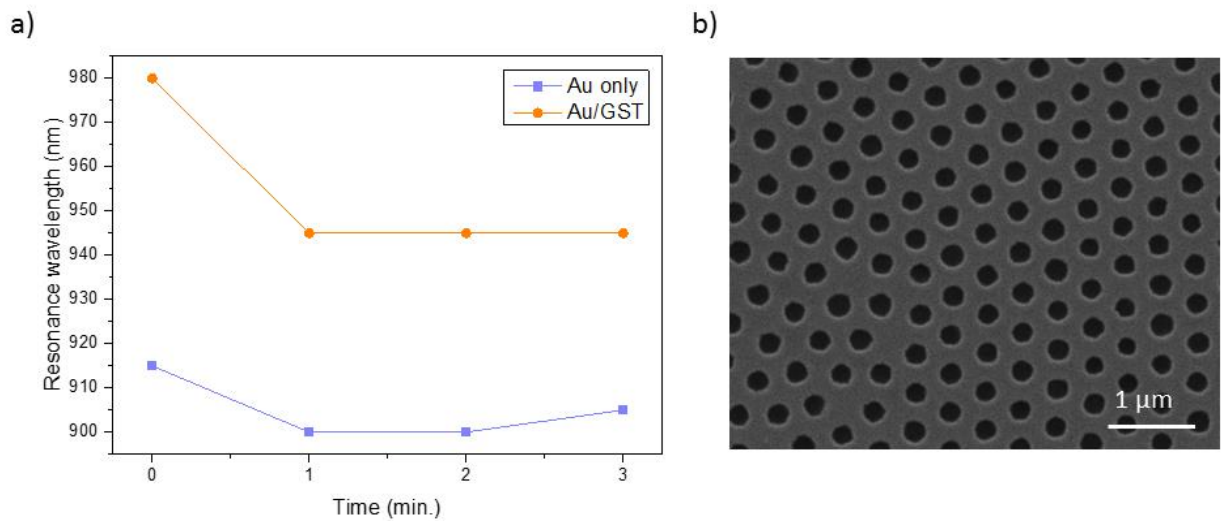

**Figure 1. a,** Time evolution of the resonance wavelength under thermal treatment at 200 ºC for the Au nanohole array with GST on top (orange) and for the bare nanohole array (purple). **b,** SEM image of a Au nanohole array covered with GST.

## II. Transmission spectra for sample B

Sample B consists of an Au nanohole array with a 20 nm GST film. GST is not present inside the holes and only covers the top flat Au surface. The fabrication procedure is identical to the one explained in the main paper, except that the PS beads are removed after deposition of GST. In this case, the GST amorphous-to-crystaline transition driven by heating produces a resonance wavelength blue shifted by 35 nm (Figure 2a), while the transmission slightly decreases from 28% to 24%, in agreement with FDTD simulations (Figure 2b). The smaller magnitude of the wavelength shift compared with sample A can be attributed to the presence of GST inside the holes in the latter, so that the material interacts with the region in which the local light intensity reaches maximum values under EOT conditions.

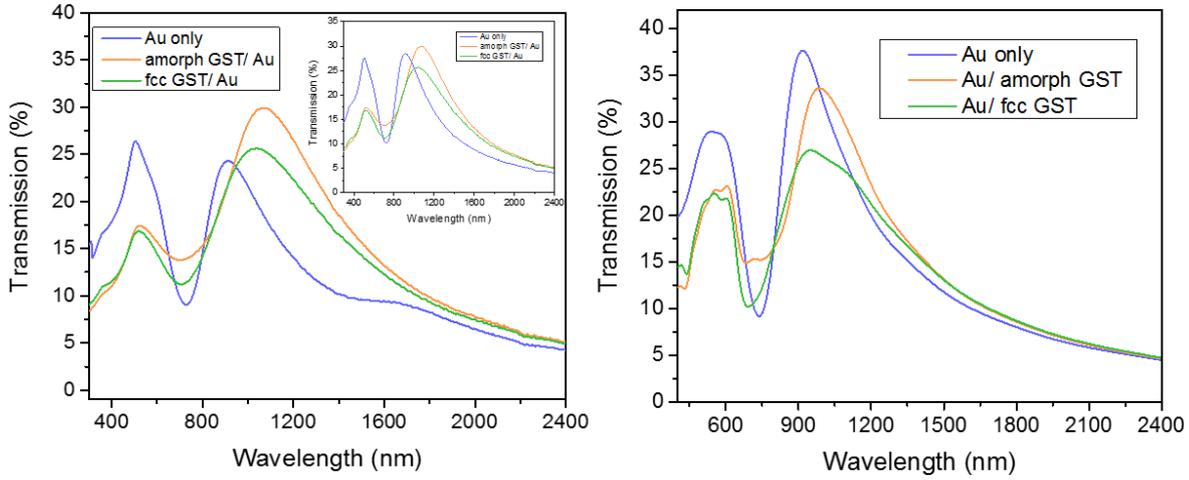

**Figure 2. a,** Measured transmission of sample B with the GST in the amorphous (orange) and crystalline (green) phases. The transmission of a bare Au nanohole array is also shown as a reference (purple). The inset shows the same measurements when the phase transition is electrically triggered applying a DC current to the device. **b,** FDTD simulation of the device for the three different cases.

In this sample we also demonstrate optical tuning by triggering the GST transition using an electric current to heat the GST layer above its crystallization temperature. Using two lateral 40 nm thick Au films as planar contact electrodes and applying a DC current (3.5 V, 1.5 A), crystallization is achieved in 20 s due to Joule heating of the Au film underneath the GST. The results shown in Figure 3c confirm that the transition to a crystalline phase produces the same optical response as that obtained by direct heating.

## III. Analytical model for periodic arrays of holes on a metallic film

We present an analytical expression for the transmission of light through a periodic array of holes in a metallic film with a phase change material (PCM) over layer. The theory is based on the dipolar scattering model of a hole drilled in metal film. In the first part of the theory effective polarizabilities of the single hole were calculated using a modal expansion of the electromagnetic field in terms of cylindrical waves assuming a perfectly conducting film. In the second part, these polarizabilities were used to calculate self-consistently the dipole moments of the holes assuming interaction between dipoles on the hole array. Finally, the calculated dipole moments were used to express the reflected and transmitted fields through scattered fields of the corresponding dipole sets on the upper and lower sides of the film.

**Polarizability of single hole**

Light transmission through hole arrays in metallic screens with finite thickness can be modelled in terms of equivalent induced electric and magnetic dipoles on either side of the film[1,2]. For simplicity, all the calculations were carried out for perfect electric conductor (PEC). Boundary conditions of Maxwell's equations permit only the presence of perpendicular electric fields and parallel magnetic fields on the surface of PEC. This means that in this model only electric dipoles perpendicular to the film surface and magnetic dipoles parallel to it will be relevant. Although the model has some limitations it becomes exact in the long wavelength region, i.e. when the hole diameter is small compared to the wavelength. Moreover in this regime the dielectric function of metals is quite large and hence metals can be treated as perfect electric conductors.

We started the theory from the calculations of the effective dipolar polarizabilities of the hole openings on both sides of the film following the so called modal expansion method given in the supplementary information of reference 3. The fields inside and outside the holes were expanded in terms of cylindrical waves and then matched at the hole openings by imposing boundary conditions of Maxwell's equations. This resulted in a system of equations for the expansion coefficients, which were then solved to find the scattered fields. Assuming that these fields were produced by the effective dipoles sitting on both sides of the metal film the effective polarizabilities of the dipoles were then extracted by comparing them to the far field of a dipole. Figure 3 shows an example of the calculated polarizabilities for a perforated PEC on top of a $SiO_2$ substrate with and without an amorphous GST layer on top (see Figure 1.b and 1.c for a schematic of the structures). As it can be seen the presence of the GST layer on top is strongly modifying the polarizability of the hole and consequently the resonance condition of the system will be different when the GST layer is present.

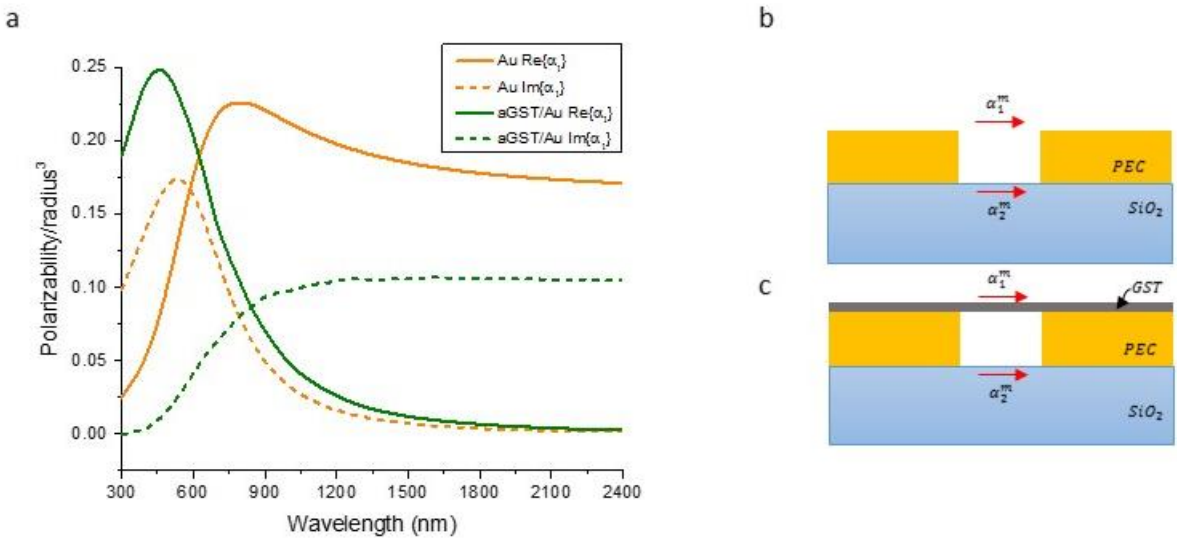

**FIGURE 3. A, CALCULATED REAL AND IMAGINARY PARTS OF MAGNETIC POLARIZABILITIES OF THE HOLE IN PEC FILM WITH (GREEN) AND WITHOUT (ORANGE) A GST LAYER ON TOP. B,C, SCHEMATICS OF THE STRUCTURES FOR WHICH THE CALCULATIONS ARE DONE.**

**Self-consistent dipoles**

For normal incidence of the field only parallel magnetic dipole moments of the system can be excited. Then using the model mentioned above for the theoretical discussion we can write a set of coupled equations for the self-consistent magnetic dipoles $m_1$ and $m_2$ at each site of the lattice corresponding to the incidence ($m_1$) and substrate ($m_2$) sides of the film[1,2]:

$$\begin{aligned} m_1 &= \alpha_{M1}(H_{ext} + G_1 m_1) + \alpha'_{M1} G_2 m_2 \\ m_2 &= \alpha'_{M2}(H_{ext} + G_1 m_1) + \alpha_{M2} G_2 m_2 \end{aligned} \quad (1)$$

Here $H_{ext}$ is the external magnetic field including also the field reflected from the film surface, $\alpha_{M1}$, $\alpha_{M2}$ and $\alpha'_{M1}$, $\alpha'_{M2}$ are the polarizabilities of the top ($\alpha_{M1}$, $\alpha'_{M1}$) and bottom ($\alpha_{M2}$, $\alpha'_{M2}$) dipoles when the field is propagating in the forward ($\alpha_{M1}$, $\alpha_{M2}$) or backward ($\alpha'_{M1}$, $\alpha'_{M2}$) direction and $G_1, G_2$ are lattice sums for the dipoles that sit on the top ($G_1$) and bottom ($G_2$) surfaces of the film.

The solution to these equations will be given as follows

$$m_1 = \frac{\alpha_1 + (\alpha'_1\alpha'_2 - \alpha_1\alpha_2)G_2}{(1-\alpha_1 G_1)(1-\alpha_2 G_2) - \alpha'_1\alpha'_2 G_1 G_2} H^{ext}$$

$$m_2 = \frac{\alpha'_2}{(1-\alpha_1 G_1)(1-\alpha_2 G_2) - \alpha'_1\alpha'_2 G_1 G_2} H^{ext}$$

(2)

The lattice sums have the form

$$G_j = \sum_{(m,n)\neq(0,0)} \mathcal{G}_j^H(\vec{R}_{mn}) e^{-i\vec{k}_\parallel \vec{R}_{mn}} \quad , j=1,2$$

(3)

Where the dyadic Green's functions $\mathcal{G}_j^H(\vec{R}_{mn})$[5] are given as

$$\mathcal{G}_j^H(\vec{R}) = -ik\left(\hat{I} \times \nabla g(R,z;\omega)\right)$$

(4)

and $g(R,\omega)$ is the scalar Green's function

$$g(r,\omega) = \frac{e^{ik_0|\vec{r}-\vec{r}'|}}{4\pi|\vec{r}-\vec{r}'|}$$

(5)

Using the Weyl's identity for spherical waves

$$\frac{e^{ik_1|\vec{r}-\vec{r}_0|}}{|\vec{r}-\vec{r}_0|} = \frac{i}{2\pi}\int \frac{d\vec{Q}}{k_{zQ1}} e^{i\vec{Q}(\vec{R}-\vec{R}_0)+ik_{zQ1}|z-z_0|}$$

(6)

and taking the appropriate derivatives inside the integral one can get all the components of the dyadic Green function. Here in particular we are only interested in $\mathcal{G}_{yy}^H$ so we will have

$$\mathcal{G}_{yy}^H(\vec{R}) = \frac{i}{2\pi}\int_0^\infty \frac{d^2\vec{Q}}{k_{zQ1}} e^{i(\vec{Q}\vec{R})} \left[\varepsilon_1 k_0^2 \left((1+r_p e^{2ik_{zQ1}z_0})\frac{Q_x^2}{Q^2} + (1-r_s e^{2ik_{zQ1}z_0})\frac{k_{zQ1}}{Q^2 k_1^2}Q_y^2\right)\right]$$

(7)

Here $r_{s,p}$ are the Fresnel coefficients for the case of a simple interface, which are given by:

$$r_s = \frac{\mu_2 k_{z1} - \mu_1 k_{z2}}{\mu_2 k_{z1} + \mu_1 k_{z2}}$$

(8)

$$r_s = \frac{\varepsilon_2 k_{z1} - \varepsilon_1 k_{z2}}{\varepsilon_2 k_{z1} + \varepsilon_1 k_{z2}}$$

(9)

The integrations in (9) can be done in two steps. The angular integration can be done using the following table[6]

$$\int \frac{d\varphi_Q}{2\pi} e^{i\vec{Q}\vec{R}} \begin{pmatrix} 1 \\ Q_x \\ Q_y \\ Q_x^2 \\ Q_y^2 \\ Q_x Q_y \end{pmatrix} = \begin{pmatrix} \jmath_0(QR) \\ \frac{iQ}{2}(\jmath_1(QR) + \jmath_{-1}(QR)) \\ \frac{Q}{2}(\jmath_1(QR) - \jmath_{-1}(QR)) \\ \frac{Q^2}{2}\left\{\jmath_0(QR) + \frac{1}{2}(\jmath_2(QR) + \jmath_{-2}(QR))\right\} \\ \frac{Q^2}{2}\left\{\jmath_0(QR) - \frac{1}{2}(\jmath_2(QR) + \jmath_{-2}(QR))\right\} \\ \frac{iQ^2}{4}(\jmath_2(QR) - \jmath_{-2}(QR)) \end{pmatrix} \qquad (10)$$

Here $\jmath_m$ is defined as $\jmath_m \equiv e^{im\phi} J_m(QR)$ and we used the following symmetry relation for the Bessel functions $J_m(QR) = (-1)^m J_{-m}(QR)$.

Angular integration leads to integrals of the form

$$I_m = \int_0^\infty f(Q)\jmath_m(QR)\, dQ \qquad (11)$$

Using the relations

$$J_m(r) = \frac{1}{2}\left(H_m^{(1)}(r) + H_m^{(2)}(r)\right)$$
$$H_m^{(2)}(-r) = -e^{in\pi} H_m^{(1)}(r) \qquad (12)$$

and upon examination of the fact that all integrand functions $f(Q)$ are even/odd for odd/even orders of the Bessel function the integrals in (11) can be brought to the form

$$I_m = I_m = \int_0^\infty f(Q)\jmath_m(QR)\, dQ = \frac{1}{2}\int_{-\infty}^\infty f(Q)\, \hbar_m^{(1)}(QR)\, dQ \qquad (13)$$

These Sommerfeld integrals are in general difficult to calculate because, for large R, Bessel functions are highly oscillatory, and this causes the integral to converge very slowly. Moreover since $f(Q) \propto r_{s,p}(Q)e^{2ik_{zQ}z}p(Q)$, the integrand has poles appearing in $r_{s,p}$ which correspond to the resonances of the system, and also there are branch cut points due to the sign ambiguity of the square root in $k_{zQ}$ at the points where $Im\{k_{zQ}\} = 0$.

To simplify the evaluation of these integrals we assumed that the main contribution to them comes from the plasmon resonance region that originates from the pole of $r_p$[3,4] (plasmon-pole approximation). In the vicinity of this resonance $r_p$ can be represented as:

$$r_p \approx \frac{2\varepsilon_1\varepsilon_2}{\varepsilon_1^2 - \varepsilon_2^2} \cdot \frac{Q_{sp}}{Q - Q_{sp}} = \frac{2k_0 B}{Q - Q_{sp}}$$
$$B = \left(\frac{\varepsilon_1\varepsilon_2}{\varepsilon_1 + \varepsilon_2}\right)^{\frac{3}{2}} / (\varepsilon_1 + \varepsilon_2) \qquad (14)$$

Where $Q_{sp} = \sqrt{\varepsilon_1\varepsilon_2/(\varepsilon_1+\varepsilon_2)}$ is the plasmon pole given as a zero of the denominator of $r_p$ in (9).

Essentially this approximation means that the holes at different sites of the lattice are interacting through long range plasmon fields. This assumption becomes exact when one considers large distances from the hole since in this region the near field contribution vanishes very rapidly.

Now using the expression (14) we can write

$$I_m(R) = Bk_0 \int_\infty^\infty \frac{g(Q)}{Q - Q_{sp}} \hbar_m^{(1)}(QR) dQ \tag{15}$$

Which by using Cauchy theorem for principal value integral will give

$$I_m(R) = 2\pi i k_0 B g(Q_{sp}) \hbar_m^{(1)}(Q_{sp}R) \tag{16}$$

Which finally leads for the $G_{yy}^H(\vec{R})$

$$G_{yy}^{HH}(R) = -\frac{\pi k_1^2 B k_0 Q_{sp}}{k_{zQ_{sp}}} \left[ \hbar_0(Q_{sp}R) + \frac{1}{2}\left(\hbar_2(Q_{sp}R) + \hbar_{-2}(Q_{sp}R)\right) \right] \tag{17}$$

Figure 2a shows a comparison of the calculated Green functions, either using a full integral or the plasmon-pole approximation. One can see that there is a little discrepancy between calculated Green function at small distances from the hole, whilst they match almost perfectly at larger distances, which clearly expresses the physical nature of the interaction between the holes.

Taking into account (17) we will have to calculate lattice sums of the form

$$S_m^{\hbar}(Q, \vec{k}_\parallel) = \sum_{(i,j) \neq (0,0)} \hbar_m^{(1)}(QR_{i,j}) e^{-i\vec{k}_\parallel \vec{R}_{i,j}} = \sum_{(i,j) \neq (0,0)} H_m^{(1)}(QR_{i,j}) e^{im\theta_{ij}} e^{-i\vec{k}_\parallel \vec{R}_{i,j}} \tag{18}$$

These sums have been evaluated numerically following the method of Kambe[7]. Figure 2b shows a plot of the lattice sum as a function of wavelength for both the top and bottom interfaces of the film. These lattice sums show characteristic divergences for wavelengths close to the lattice period. These divergences lead to lattice resonances which manifest in the transmission and reflection spectra. Moreover the lattice sum in the substrate region ($G_2$) shows an order of magnitude larger value than that in the superstrate region ($G_1$). This is due to the stronger localized plasmon modes on SiO$_2$/Au interface compared to the ones present in an air/Au interface. The stronger the localization the more intense is the plasmon field and thus the dipole-dipole interaction is stronger.

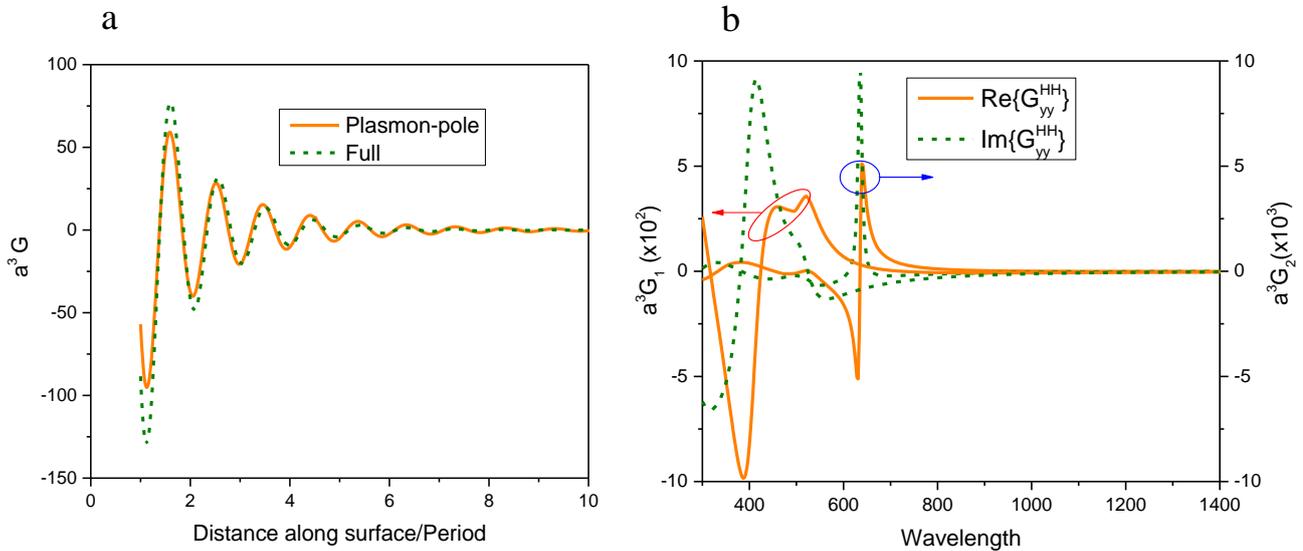

Once the dipole moment in the substrate side of the film ($m_2$) is known we can find the transmission coefficient of the system summing up the contribution from all dipoles sitting in that interface. The electric field radiated by a magnetic dipole $\vec{m}$ is given as

**Figure 4. a, b,** Calculated lattice sums under the plasmon-pole approximation compared to the full integral calculation.

$$\vec{E}_{i,scat}(\vec{r}) = [-ik_0 \vec{m} \times \vec{\nabla}] \frac{e^{ik_2|\vec{r}-\vec{r}_i|}}{|\vec{r}-\vec{r}_i|} \tag{19}$$

Taking into account that in our case $\vec{m} = m_y \hat{e}_y = m_2 \hat{e}_y$ and taking the superposition of all fields a straightforward calculation gives the total scattered field in the far field:

$$\vec{E}_{far}^{tot} = \frac{2\pi i}{A} \frac{km_2(k_{z2}\hat{e}_x - k_x\hat{e}_z)}{k_{2z}} e^{i(\vec{k}_\parallel \vec{R} + k_{2z}|z-z_0|)} \tag{20}$$

which for normal incidence immediately leads for the transmission coefficient

$$t = \frac{2\pi i k_0 m_2}{A}(1 - r_{23}) \tag{21}$$

These are the formulas used to calculate the transmission curves in main text of the paper.